\documentclass[10pt,conference]{IEEEtran}
\IEEEoverridecommandlockouts
\usepackage{cite}
\usepackage{amsmath,amssymb,amsfonts}
\usepackage{algorithmic}
\usepackage{graphicx}
\usepackage{textcomp}
\usepackage{enumitem}
\usepackage{xcolor}
\def\BibTeX{{\rm B\kern-.05em{\sc i\kern-.025em b}\kern-.08em
    T\kern-.1667em\lower.7ex\hbox{E}\kern-.125emX}}

\usepackage{pifont}
\newcommand{\cmark}{\ding{51}}  % ✓
\newcommand{\xmark}{\ding{55}}  % ✗
\usepackage{multirow}
\usepackage{booktabs}
\usepackage{tikz}
\usetikzlibrary{positioning, shapes, calc}
\usepackage{url}
\usepackage{graphicx}
\usepackage{array}
\usepackage{subcaption}
\usepackage{float}

\begin{document}

\title{Key Considerations for Auto-Scaling: \\Lessons from Benchmark Microservices\\}
% Auto-Scaling Challenges in Microservice Migration: Issues and Solutions
% Enabling Effective Auto-Scaling in Microservices: Key Considerations for Monolithic Migration
% From Monolith to Microservices: Addressing Auto-Scaling Limitations in Migrated Applications
% Improving Auto-Scaling Readiness in Microservice Migration: Architectural and Operational Insights
% A Framework for Auto-Scaling-Aware Migration from Monolithic to Microservice Architectures

%\author{\IEEEauthorblockN{Majid Dashtbani and Ladan Tahvildari}
%\IEEEauthorblockA{\textit{Department of Electrical and Computer Engineering} \\
%\textit{University of Waterloo}\\
%Waterloo, Canada \\
%\{mdashtbani,ladan.tahvildari\}@uwaterloo.ca}
%}

\author{\IEEEauthorblockN{Majid Dashtbani and Ladan Tahvildari}
\IEEEauthorblockA{\textit{Department of Electrical and Computer Engineering} \\
\textit{University of Waterloo}\\
Waterloo, Canada \\
\{majid.dashtbani,ladan.tahvildari\}@uwaterloo.ca}
}

\maketitle

\begin{abstract}
Microservices have become the dominant architectural paradigm for building scalable and modular cloud-native systems. However, achieving effective auto-scaling in such systems remains a non-trivial challenge, as it depends not only on advanced scaling techniques but also on sound design, implementation, and deployment practices. Yet, these foundational aspects are often overlooked in existing benchmarks, making it difficult to evaluate autoscaling methods under realistic conditions. In this paper, we identify a set of practical auto-scaling considerations by applying several state-of-the-art autoscaling methods to widely used microservice benchmarks. To structure these findings, we classify the issues based on when they arise during the software lifecycle: \textit{Architecture}, \textit{Implementation}, and \textit{Deployment}. The \textit{Architecture} phase covers high-level decisions such as service decomposition and inter-service dependencies. The \textit{Implementation} phase includes aspects like initialization overhead, metrics instrumentation, and error propagation. The \textit{Deployment} phase focuses on runtime configurations such as resource limits and health checks. We validate these considerations using the Sock-Shop benchmark and evaluate diverse auto-scaling strategies—including threshold-based, control-theoretic, learning-based, black-box optimization, and dependency-aware approaches. Our findings show that overlooking key lifecycle concerns can degrade autoscaler performance, while addressing them leads to more stable and efficient scaling. These results underscore the importance of lifecycle-aware engineering for unlocking the full potential of auto-scaling in microservice-based systems.
\end{abstract}


\section{Introduction}\label{sec:introduction}
Microservices have emerged as the dominant architectural paradigm for building scalable, modular, and cloud-native systems~\cite{microdecomposition,b2}. By decomposing applications into independently deployable services, microservices enable rapid iteration, resilience, and horizontal scaling—properties that are especially well-aligned with the elastic nature of cloud computing.

A fundamental feature of microservices is their elasticity—the ability to dynamically provision and release computing resources based on real-time workload fluctuations~\cite{b3}. This capability is beneficial in cloud environments, where scaling resources in response to demand is essential for maintaining performance and optimizing cost. However, while microservice architectures inherently support elasticity, realizing its benefits in practice requires not only scalable infrastructure but also the adoption of effective strategies for dynamic resource management~\cite{Dashtbani2025}.

Although cloud platforms offer elastic infrastructure, effective elasticity is not guaranteed. Responsiveness and stability under variable workloads depend on automated resource adjustment mechanisms—commonly referred to as auto-scaling.

While prominent industry players ~\cite{WhyMicro} have leveraged microservice elasticity to improve quality of service (QoS) and reduce operational costs, their internal auto-scaling strategies remain largely undisclosed. In contrast, the academic community has proposed a wide range of auto-scaling techniques—ranging from rule-based and control-theoretic models \cite{Dashtbani2025} to machine learning approaches \cite{DCScaler,PBScaler}—to manage elasticity in microservice systems. However, these academic efforts often rely on simplified microservice benchmarks and simulated workloads, which may fail to capture the complexities of production systems. This gap raises important questions about the practical applicability of proposed methods.

To address this gap, we conduct a study of auto-scaling behaviors in widely used microservice benchmarks. By applying a range of auto-scaling methods to real workloads, we uncover recurring issues that affect scaling accuracy, efficiency, and stability. We organize these issues into three key phases of the microservice lifecycle: \textit{Architecture}, \textit{Implementation}, and \textit{Deployment}. These phases capture decisions about service chaining and dependencies, runtime instrumentation and error propagation, and configuration parameters such as resource quotas and readiness probes.

This paper provides practical insights into how these lifecycle decisions impact auto-scaler performance, and shows that overlooking key considerations in any phase can significantly degrade elasticity. We validate our findings through empirical evaluation of six representative auto-scaling strategies on the Sock-Shop benchmark\footnote{https://github.com/microservices-demo/microservices-demo}, highlighting how better lifecycle engineering enables more effective resource management.

The remainder of the paper is organized as follows: Section~\ref{sec:motivation} introduces the motivation. Sections~\ref{sec:challenges}–\ref{sec:results} cover the identified challenges, experimental setup, and results. Section~\ref{sec:lessons} outlines lessons learned, Section~\ref{sec:background} reviews related work, and Section~\ref{sec:conclusion} concludes.  

\begin{table*}[t]
  \centering
  \caption{Overview of microservice benchmarks and observed auto-scaling issues (\cmark: Yes, \xmark: No)}
  \label{tab:challenges-matrix}
  \setlength{\tabcolsep}{5pt}
  \renewcommand{\arraystretch}{1.2}
  \begin{tabular}{@{}lcl
    *{6}{>{\centering\arraybackslash}p{1.0cm}}@{}}
    \toprule
    \multicolumn{3}{c}{} 
    & \multicolumn{1}{c}{\textbf{Scalability}} 
    & \multicolumn{4}{c}{\textbf{Observability}} 
    & \multicolumn{1}{c}{\textbf{Security}} \\

    % Insert the reference below the header, two lines down
    \multicolumn{3}{c}{} 
    & \multicolumn{1}{c}{\scriptsize(Sec. \ref{sec:challenges}.A)} 
    & \multicolumn{4}{c}{\scriptsize(Sec. \ref{sec:challenges}.B)} 
    & \multicolumn{1}{c}{\scriptsize(Sec. \ref{sec:challenges}.C)} \\
    
    \cmidrule(lr){4-4} \cmidrule(lr){5-8} \cmidrule(lr){9-9}

    \textbf{Benchmarks} & \textbf{Microservices} & \textbf{Languages} 
    & \rotatebox[origin=c]{90}{\parbox{1.3cm}{\centering Heavy\\Services}} 
    & \rotatebox[origin=c]{90}{Monitoring} 
    & \rotatebox[origin=c]{90}{Readiness}
    & \rotatebox[origin=c]{90}{\parbox{1.3cm}{\centering Failure\\Visibility}} 
    & \rotatebox[origin=c]{90}{Dependencies} 
    & \rotatebox[origin=c]{90}{\parbox{1.3cm}{\centering Resource\\Governance}} \\
    \midrule

    Bookinfo        
    & 4  
    & \scriptsize Java, Python, Node.js, Ruby
    & \cmark & \xmark & \xmark & \xmark & \xmark & \xmark \\

    Online Boutique  
    & 11 
    & \scriptsize Java, Python, Node.js, Go, C\#
    & \cmark & \xmark & \cmark & \xmark & \xmark & \cmark \\

    Sock Shop       
    & 13 
    & \scriptsize Java, Python, Node.js, Go 
    & \cmark & \cmark & \cmark & \cmark & \cmark & \xmark \\

    TrainTicket    
    & 41 
    & \scriptsize Java, Python, Node.js, Go, C\#
    & \cmark & \cmark & \cmark & \cmark & \xmark & \cmark \\

    \bottomrule
  \end{tabular}
\end{table*}

\section{Motivation}\label{sec:motivation}
The growing demand for scalable and reliable online services has driven widespread adoption of microservice architectures. These architectures enable the flexible, independent scaling of service components. Auto-scaling complements this by dynamically adjusting resources based on workload fluctuations, thereby improving performance and reducing operational costs.

Netflix exemplifies this shift. By migrating from a monolithic to microservices\footnote{https://roshancloudarchitect.me/understanding-netflixs-microservices-architecture-a-cloud-architect-s-perspective-5c345f0a70af}, Netflix improved its scalability, resilience, and development agility \footnote{https://netflixtechblog.com/rebuilding-netflix-video-processing-pipeline-with-microservices-4e5e6310e359}. Auto-scaling has played a critical role in allowing the platform to support over 200 million users while optimizing infrastructure usage\footnote{https://aws.plainenglish.io/how-netflix-hyperscales-aws-inside-its-200m-user-infrastructure-with-auto-scaling-chaos-80b3ff9f1ede}.

\noindent\textbf{Microservice Benchmarks.} Despite growing industry adoption, the internal auto-scaling strategies used by leading organizations remain largely proprietary. As a result, academic research often relies on publicly available microservice benchmarks—such as Bookinfo\footnote{https://github.com/istio/istio/tree/master/samples/bookinfo}, Online Boutique\footnote{https://github.com/GoogleCloudPlatform/microservices-demo}, Sock-Shop, and TrainTicket\footnote{https://github.com/FudanSELab/train-ticket}—to evaluate new auto-scaling techniques. Table~\ref{tab:challenges-matrix} summarizes these systems in terms of microservice count, language diversity, and observable auto-scaling issues.

\noindent\textbf{Problem Definition.} While numerous advanced autoscaling methods have been proposed, their practical evaluation is hindered by the limitations of widely used microservice benchmarks. These demo systems often omit key features—such as complete call graphs, failure propagation, or resource governance—that are essential to support accurate and effective scaling. This means, \textit{even well-designed autoscalers underperform unless these gaps are addressed}.

\noindent\textbf{Potential Solution.} Our research work aims not only to show that resolving these issues improves autoscaler performance, but also to identify a set of actionable design considerations that can guide the development of microservices more compatible with intelligent autoscaling.

To better understand and address these pitfalls, we decompose the existing issues into three high-level challenges:

\begin{itemize}[leftmargin=0.3cm]
  \item \textbf{Scalability}: Refers to the system’s ability to elastically respond to load, especially for heavy services.
  \item \textbf{Observability}: Refers to gaps in visibility—metrics, probes, or tracing—that may mislead the autoscaler.
  \item \textbf{Security}: Refers to risks of unbounded resource usage and denial-of-service attacks when limits or quotas are missing.
\end{itemize}

Table~\ref{tab:challenges-matrix} summarizes which of these considerations were observed in four widely used microservice benchmarks. In the next section, we describe these considerations in detail, highlighting how and where they manifest in practice.

\section{Auto-Scaling challenges in Practice}\label{sec:challenges}
To evaluate academic auto-scaling methods, we deployed four benchmarks on a Kubernetes\footnote{https://kubernetes.io}
 cluster using a high-performance server; the benchmarks are described in Table~\ref{tab:challenges-matrix}. During our experiments, we found that these benchmarks were not fully compatible with state-of-the-art auto-scaling methods, revealing several practical challenges. Among them, Sock-Shop exhibited the highest number of scaling-related issues, making it our primary case study. As an online shopping application, Sock-Shop includes several microservices; we focused on the \texttt{/login} endpoint, which involves a representative service chain: \texttt{Front-end}~$\rightarrow$~\texttt{User}~$\rightarrow$~\texttt{Carts}. In the following, we present each challenge as a gap, labeled using the prefix ``G''.

\subsection{Scalability}\label{sec:Scalability}
Scalability is essential for microservices to adapt to changing workloads. However, we observed key issues that limit effective scale-out, especially in services with high startup costs or resource demands. This section outlines such considerations, starting with heavy services.

\vspace{0.2em}
\noindent\textbf{Heavy Services.}
Certain microservices consume significant resources during startup or scale-out operations, posing unique challenges to autoscalers. These heavy services can trigger misleading metric signals or strain cluster capacity when scaled aggressively. We divide the resulting challenges into two main types: service initialization overhead and scale-out resource contention.

\begin{figure*}[ht]
  \centering

  \begin{subfigure}[b]{0.49\textwidth}
    \includegraphics[width=\textwidth]{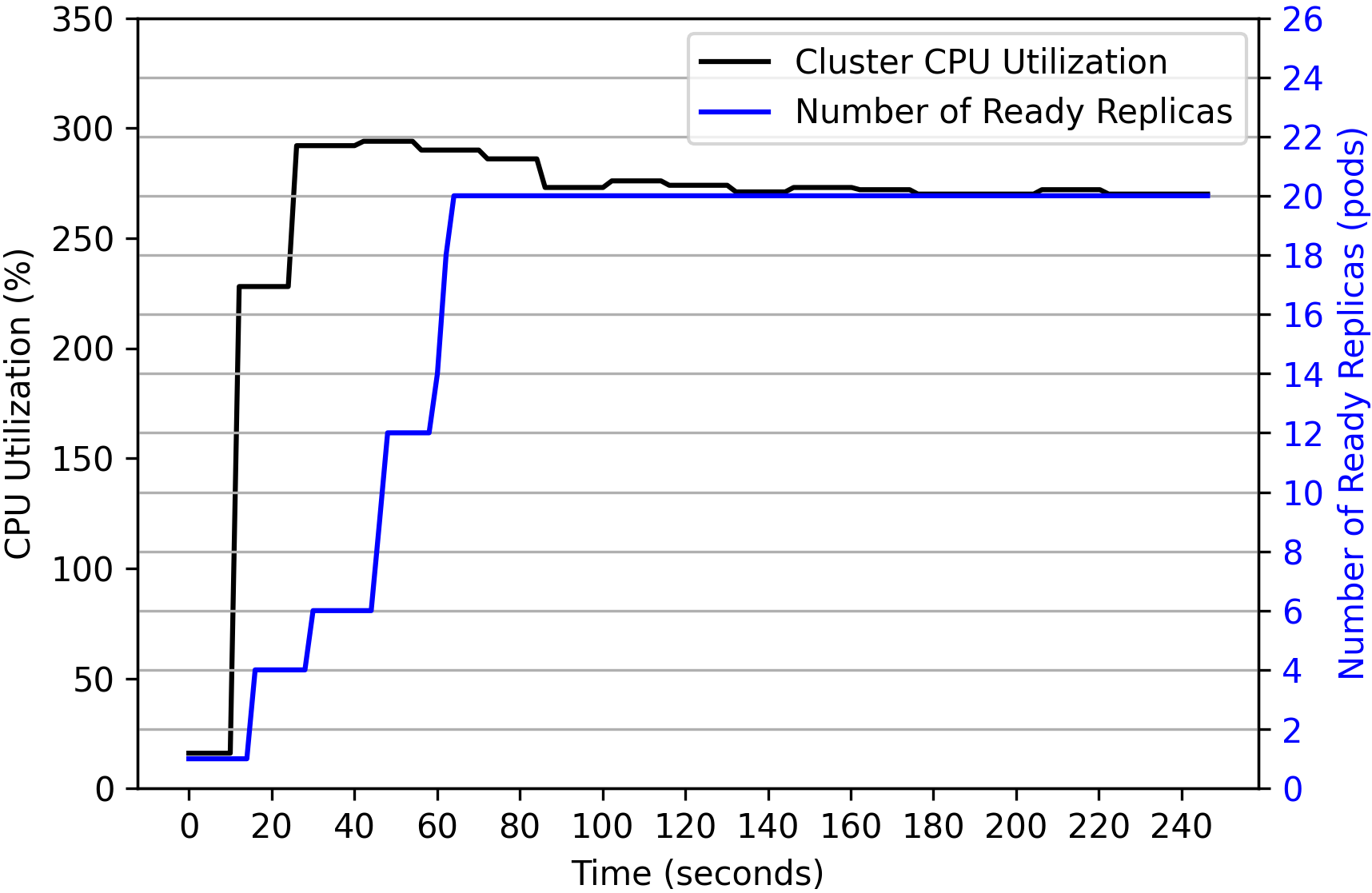}
    \caption{Replica count increase due to CPU spikes during initialization.}
    \label{fig:init-spike-replicas}
  \end{subfigure}
  \hfill
  \begin{subfigure}[b]{0.49\textwidth}
    \includegraphics[width=\textwidth]{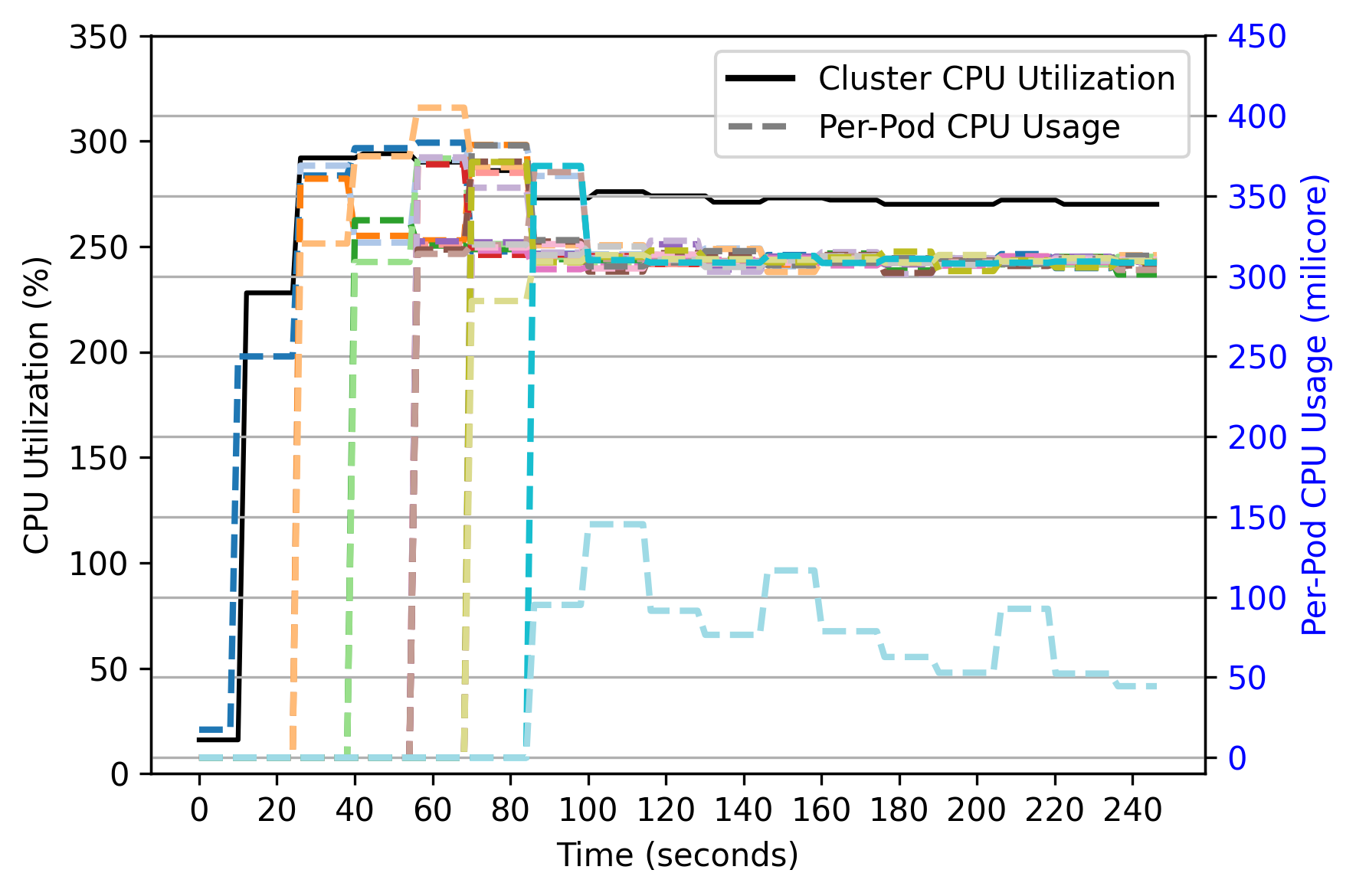}
    \caption{Per-pod CPU usage shows high variance during startup.}
    \label{fig:init-spike-perpod}
  \end{subfigure}
  \hfill

  \caption{Impact of heavy services on auto-scaling performance. 
  (a) and (b) demonstrate how service initialization overhead triggers false-positive scale-outs under KHPA - Max pods: 20; Resource configuration: Requests: 100m, Limits: 300m. 
  }
  \label{fig:heavy-service-effects}
  \vspace{-1.0em}
\end{figure*}

\noindent\textbullet\ \textit{G1: Service Initialization Overhead:}
During our experiments with the Sock-Shop benchmark, we applied Kubernetes Horizontal Pod Autoscaler (KHPA)\footnote{https://kubernetes.io/docs/tasks/run-application/horizontal-pod-autoscale/} to the login service chain with a CPU threshold of 50\%. We observed anomalous behavior: the autoscaler triggered scale-out events even when no external workload was present.

Further investigation revealed that the \texttt{Carts} service, implemented as a Java Spring application, exhibits high CPU usage during JVM startup. KHPA interpreted this temporary spike as load, launching unnecessary replicas. These new replicas also exhibited the same behavior, creating a loop of runaway scaling. 

This issue is illustrated in Fig.~\ref{fig:init-spike-replicas}, where the total replica count increases even in the absence of real user load. Fig.~\ref{fig:init-spike-perpod} further shows the per-pod CPU usage, highlighting the variability and intensity of startup overheads across individual replicas.  

To mitigate this issue, heavy startup logic should be encapsulated in an \texttt{initContainer} with separate resource quotas and lifecycle. This ensures the main container’s metrics reflect steady-state behavior only, preventing autoscaler reactions to transient initialization spikes. However, most academic autoscalers and benchmarks lack such separation, causing inflated resource measurements during cold starts. An additional mitigation—discussed next—involves using readiness and liveness probes to suppress misleading startup metrics.

\noindent\textbullet\ \textit{G2: Scale-Out Resource Contention:}
Even with proper probes and init-containers, bulk scaling of heavy services like \texttt{Carts} introduces a second-order problem. Simultaneously launching multiple replicas with high startup demands (e.g., loading frameworks or caches) can consume significant CPU and memory, leading to resource contention with co-located microservices and degrading cluster-level performance. 

\begin{figure}[H]
  \centering
  \includegraphics[width=0.48\textwidth]{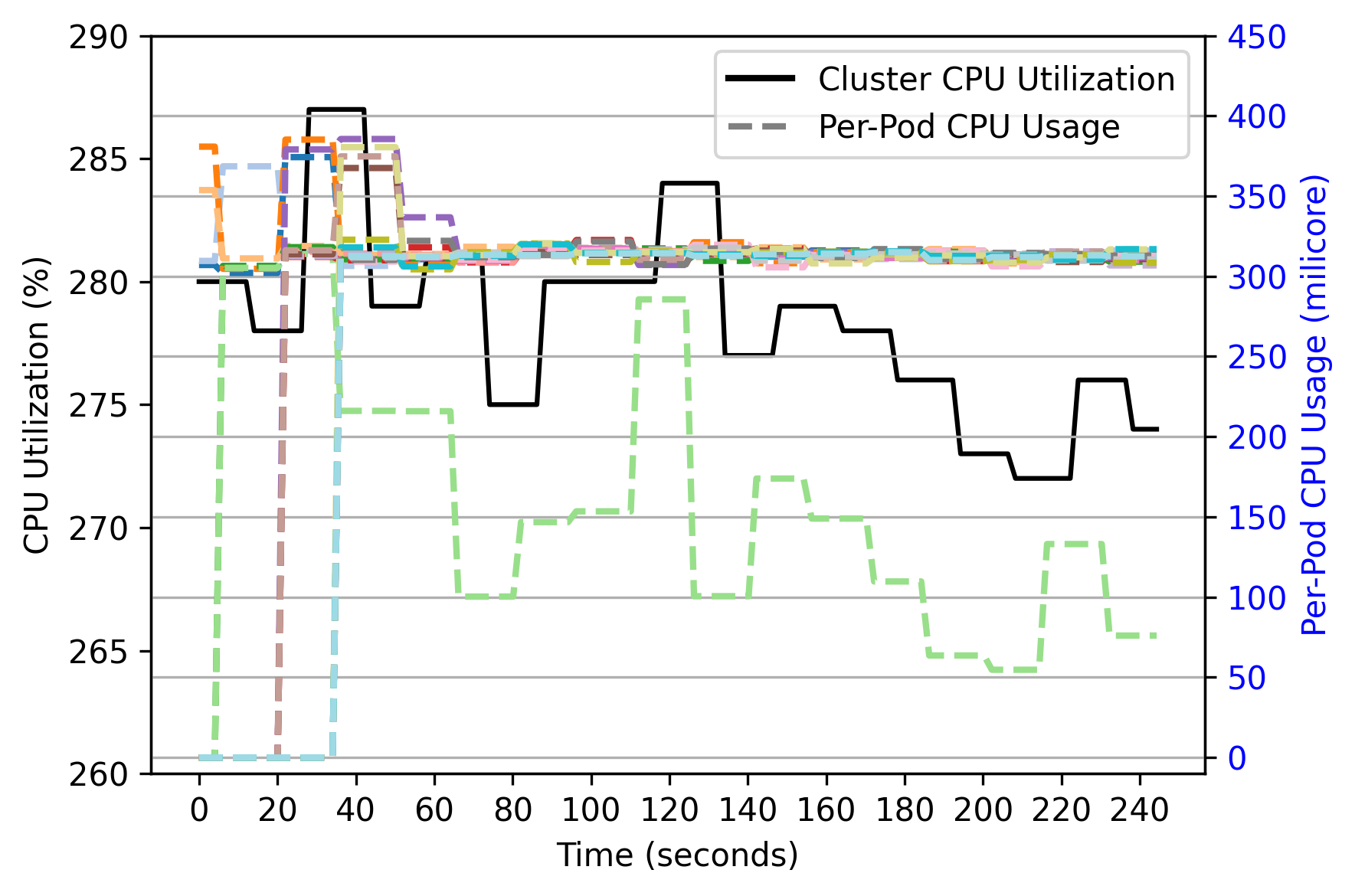}
  \caption{CPU contention - parallel startup of multiple replicas.}
  \label{fig:scale-contention}
\end{figure}

This scenario is visualized in Fig.~\ref{fig:scale-contention}, where CPU usage across pods overlaps significantly during simultaneous startup, stressing shared resources and delaying service stabilization. To mitigate this, autoscalers should apply rate limits to pod creation, and administrators must define resource quotas to preserve cluster balance during scale-out events.

\vspace{-0.5em}

\subsection{Observability}
Effective auto-scaling depends on accurate visibility into service health, performance, and inter-service dependencies. However, many demo microservices lack sufficient observability, making it difficult for autoscalers to react appropriately. In this section, we group our observations into four areas aligned with Table~\ref{tab:challenges-matrix}: \emph{Monitoring}, \emph{Readiness}, \emph{Failure Visibility}, and \emph{Dependencies}.

\vspace{0.2em}
\noindent\textbf{Monitoring.}
Some services lack essential monitoring metrics, limiting the autoscaler's ability to make informed decisions. Without data such as latency or error rates, scaling becomes reactive, delayed, and potentially inaccurate.

\noindent\textbullet\ \textit{G3: Missing Application-Level Metrics:}
To prevent KHPA from scaling during JVM warm-up, we attempted to configure boot and health checks for the \texttt{Carts} service. This required exposing an HTTP status path to report health information to Kubernetes, which was not enabled by default. We had to modify the service’s Java options to expose this internal status.

This highlights a deeper issue: the absence of application-level observability. Effective auto-scaling requires both low-level system metrics (e.g., CPU, memory, I/O) and high-level application metrics (e.g., latency, error rate, request queue length). Modern auto-scaling methods, such as those in\cite{DeepScaler,MultiMetricTh}, use multi-metric models—often driven by machine learning—to capture temporal and causal interactions between services. Supporting these methods requires exposing metrics across multiple layers of granularity. Services should be instrumented to export both application and infrastructure data and integrate seamlessly with telemetry pipelines and service mesh tools (e.g., Istio\footnote{https://istio.io}) for complete observability.

\vspace{0.7em}
\noindent\textbf{Readiness.}
Readiness probes indicate whether a service is prepared to handle incoming requests. Without them, autoscalers may count unready pods, leading to inaccurate scaling decisions and degraded performance.

\noindent\textbullet\ \textit{G4: Missing Readiness/Liveness configuration:}
Probes play a critical role in suppressing autoscaler responses to transient conditions like cold starts. In the \texttt{Carts} service, we used a readiness probe that marked the pod as ready only after the JVM was fully initialized. This ensured KHPA didn’t count the pod until it could serve real traffic. Similarly, a liveness probe with startup delay helped avoid unnecessary restarts triggered by slow boot times. Although helpful, probe configuration remains non-trivial and can itself become a source of instability—especially when infrastructure assumptions are not portable across environments.

\noindent\textbullet\ \textit{G5: Misconfigured Readiness/Liveness Probe:}
Improper probe configuration can cause severe disruptions. In our experiments with the TrainTicket benchmark, pods repeatedly restarted due to aggressive timeout settings. These thresholds were likely designed for high-performance infrastructure. When deployed on modest clusters, the services could not boot fast enough to meet the configured readiness deadline. Kubernetes interpreted this as a failure and restarted the pod, creating an availability loop. This emphasizes the need for probe configurations that reflect both the platform and resource constraints.

\vspace{0.2em}
\noindent\textbf{Failure Visibility.}  
Effective auto-scaling relies on accurate detection of failures across service dependencies. When microservices suppress or misreport errors, autoscalers and monitoring systems may overlook critical degradations, delaying recovery actions and compromising system reliability.

\noindent\textbullet\ \textit{G6: Error Masking in Service Chains:}
After resolving initialization issues with the \texttt{Carts} service, we evaluated end-to-end behavior by sending login traffic through the Sock-Shop benchmark. We used the 90th percentile (P90) latency of the \texttt{Front-end} service as a soft Service-Level Objective (SLO). While some test runs showed acceptable latency, others reported excellent response times even with downstream failures.

\begin{figure}[ht]
  \centering
  \includegraphics[width=0.48\textwidth]{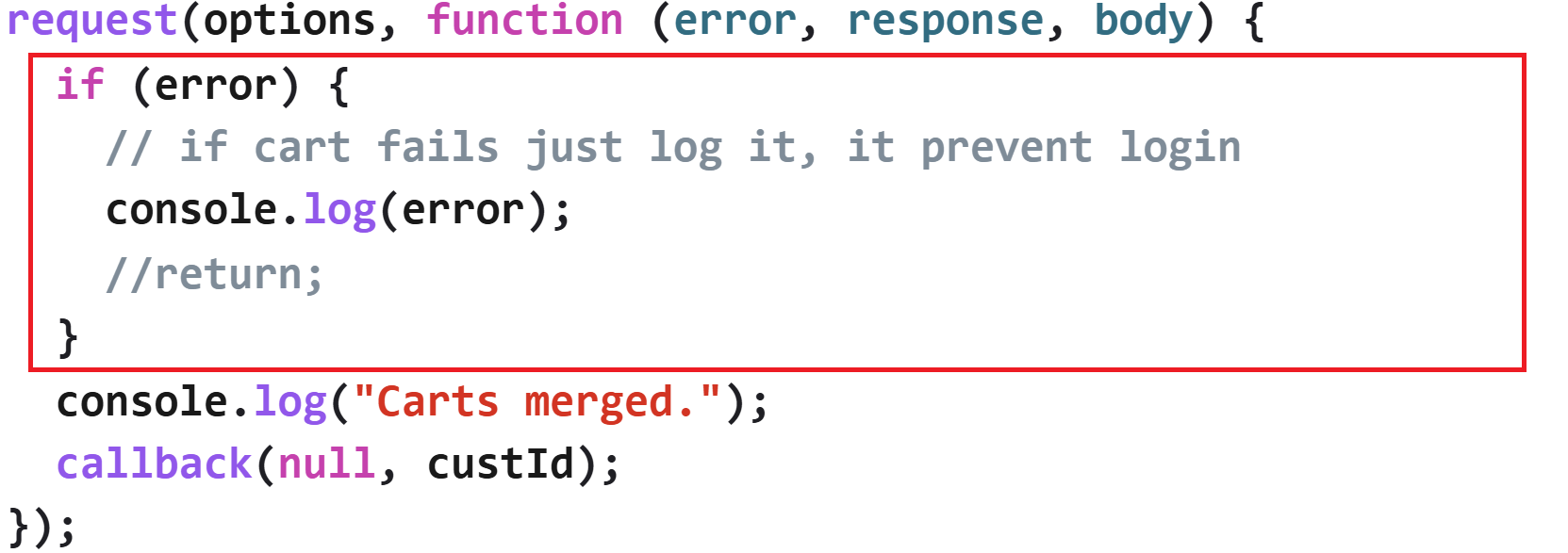}
  \caption{Front-end code masks \texttt{Carts} failure with HTTP 200.}
  \label{fig:cartsloginerrorcode}
  \vspace{-0.5em}
\end{figure}

Closer inspection revealed that the \texttt{Carts} service was silently failing. As shown in Fig.~\ref{fig:cartsloginerrorcode}, the \texttt{Front-end} service caught exceptions from \texttt{Carts} and returned an HTTP 200 success response—despite the shopping cart being unavailable. This behavior masked the failure from both users and the auto-scaler, preventing any corrective scaling action.

\begin{figure}[htpb]
  \centering
  \includegraphics[width=0.49\textwidth]{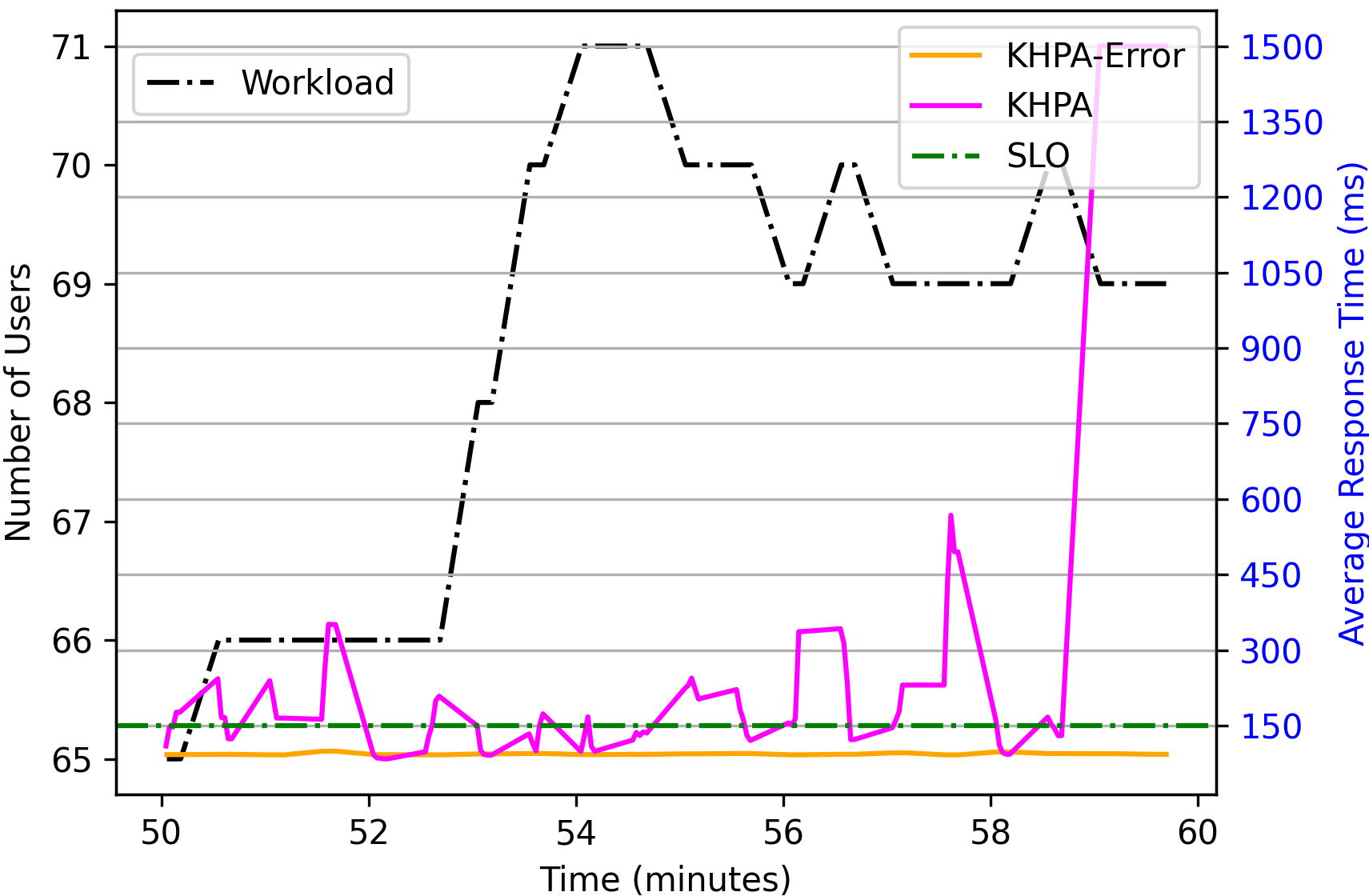}
  \caption{Error masking lowers response time, misleading the autoscaler.}
  \label{fig:cartsresponsecompare}
  \vspace{-0.5em}
\end{figure}

To quantify the impact, we compared two experiments: one in which \texttt{Carts} failures were surfaced properly (KHPA), and one in which they were masked (KHPA-Error). Fig.~\ref{fig:cartsresponsecompare} shows that in HPA-Error, the response time remains artificially low, since the failed service returns immediately. In contrast, KHPA reflects the actual latency caused by a stressed \texttt{Carts} service, enabling the auto-scaler to trigger appropriate scaling actions.

This illustrates how exception masking undermines auto-scaler accuracy and reliability. Services must propagate failures using correct HTTP status codes and emit metrics that reflect degraded downstream behavior.

\noindent\textbullet\ \textit{G7: Lack of Downstream Error Metrics:}
To detect service degradation beyond status codes, applications should emit explicit error metrics such as \texttt{downstream\_error}, \texttt{retry\_failure}, or \texttt{dependency\_unavailable}. These metrics enable auto-scalers and operators to make scaling or routing decisions based on deeper service health semantics. Integration with tracing systems and service meshes like Istio can further enhance observability, enabling mechanisms such as circuit breaking, retry limits, and granular telemetry to expose latent request-path failures.

\vspace{0.2em}
\noindent\textbf{Dependencies.}
Accurate service dependency modeling is essential for auto-scalers that use call graphs to detect bottlenecks and guide scaling. However, poor instrumentation, limited observability, or architectural choices can obscure these relationships, leading to misleading graphs and suboptimal scaling decisions.

\noindent\textbullet\ \textit{G8: Incomplete or Misleading Call Graphs:}
Beyond KHPA, we experimented with more advanced auto-scaling frameworks such as PBScaler and DeepScaler. These methods often rely on service-level dependency graphs—either pre-constructed or dynamically inferred—to identify bottlenecks and inform scaling. A key architectural insight from our experiments is that the accuracy and utility of these graphs are highly sensitive to how service dependencies are implemented and instrumented. In modern cloud-native stacks, service meshes like Istio can generate runtime call graphs by observing inter-service communications. While promising, these graphs can still be incomplete or misleading if architectural patterns obscure causality or observability hooks are missing.

\begin{figure}[ht]
  \centering
  \includegraphics[width=0.48\textwidth]{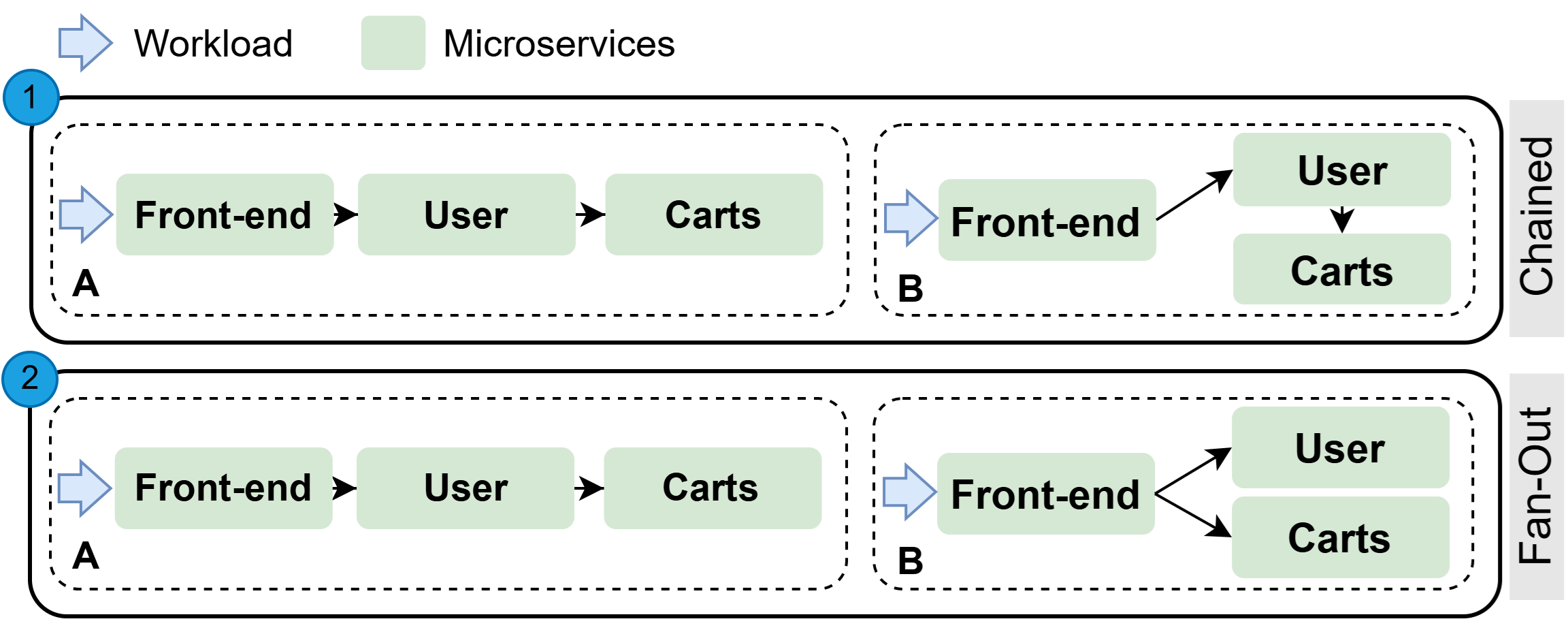}
  \caption{“/login” chain: (A) Business logic view; (B) Runtime invocation view.}
  \label{fig:loginCallsAbstract}
\end{figure}

To illustrate this challenge, consider the “/login” service chain in our Sock-Shop case study, which involves the \texttt{Front-end}, \texttt{User}, and \texttt{Carts} services. Fig.~\ref{fig:loginCallsAbstract} contrasts two invocation patterns for this chain. In the left pane (A), the Business Logic View presents the high-level intent: \texttt{User} authentication followed by \texttt{Carts} retrieval. However, it abstracts away the runtime call flow. The right pane (B), the Service Invocation View, depicts the actual sequence of inter-service calls.

\begin{figure}[ht]
  \centering
  \includegraphics[width=0.48\textwidth]{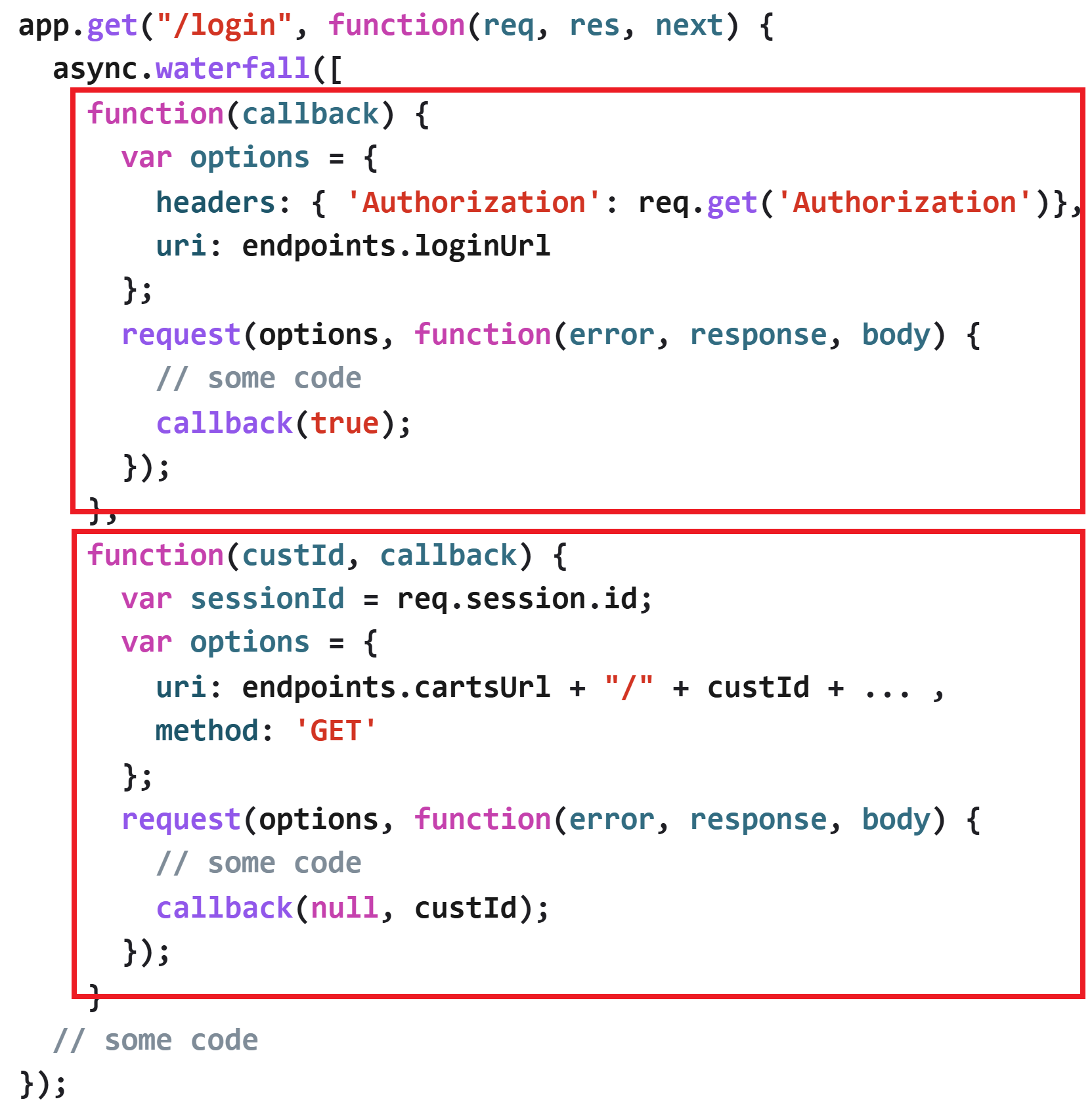}
  \caption{Front-end “/login” code showing Fan-Out calls to User and Carts.}
  \label{fig:loginCallsSourceCode}
  \vspace{-1.0em}
\end{figure}

\begin{figure}[ht]
  \centering
  \includegraphics[width=0.48\textwidth]{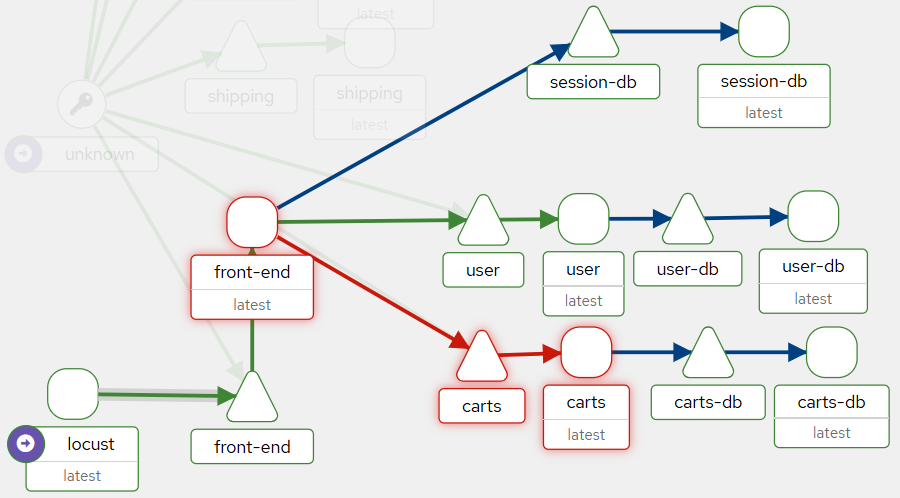}
  \caption{Call graph generated by Istio for the Sock-Shop “/login” flow.}
  \label{fig:loginCallISTIO}
\end{figure}

%\vspace{-1em}
Two patterns emerge: (1) \emph{Chained Invocation}, where \texttt{Front-end} calls \texttt{User}, which then calls \texttt{Carts}; and (2) \emph{Fan-Out Invocation}, where \texttt{Front-end} directly calls both. As shown in Fig.~\ref{fig:loginCallsSourceCode}, Sock-Shop adopts the Fan-Out pattern, also reflected in the Istio-generated graph (Fig.~\ref{fig:loginCallISTIO}).

%\vspace{-0.7em}

\begin{figure}[H]
  \centering
  \includegraphics[width=0.49\textwidth]{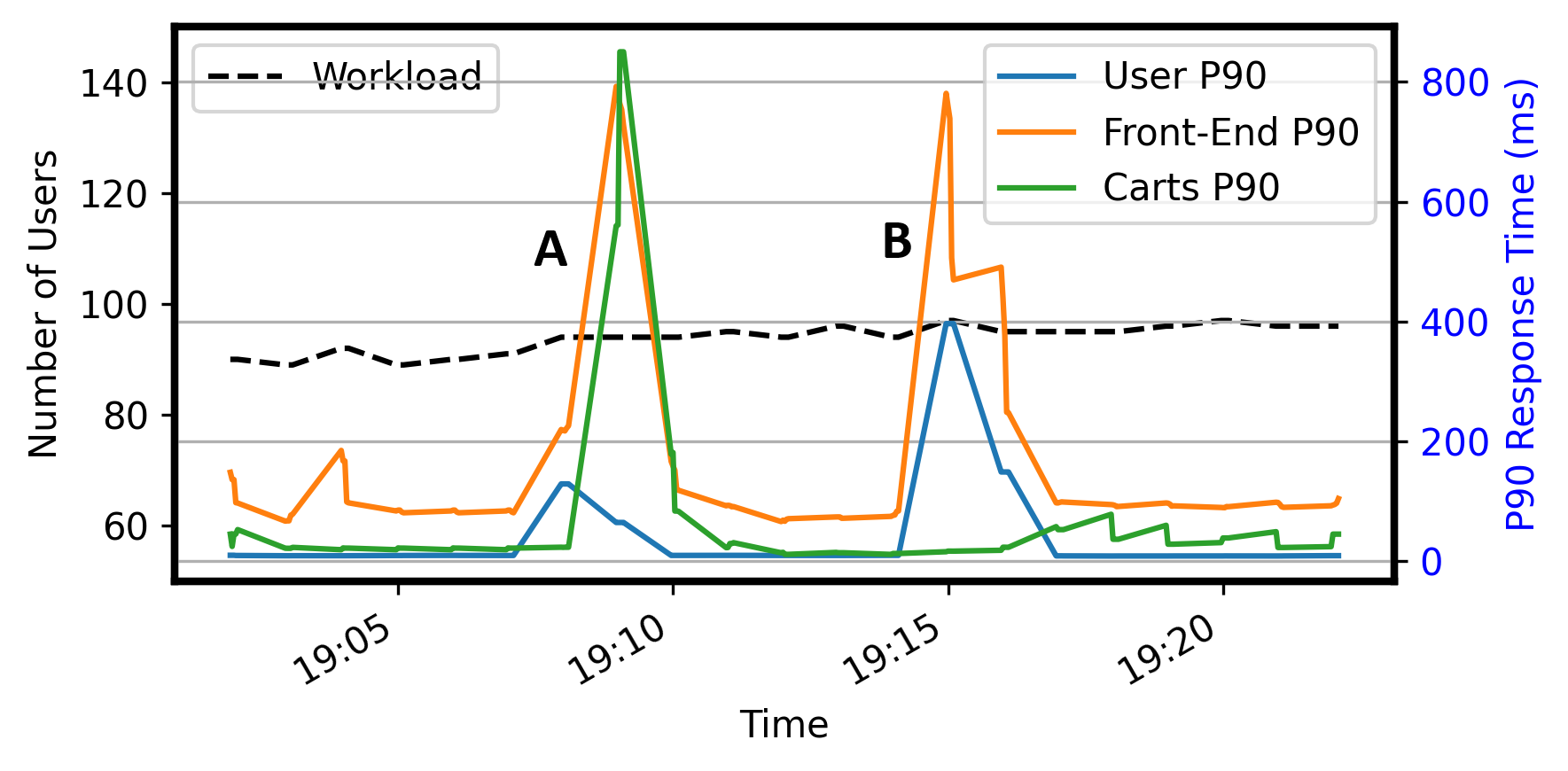}
  \caption{P90 latency showing (A) \texttt{Carts} and (B) \texttt{User} as bottlenecks, causing SLO violation in \texttt{Front-end}}
  \label{fig:bottleneckP90}
\end{figure}

While functionally equivalent, these designs differ in observability. The Fan-Out model breaks the causal link between \texttt{User} and \texttt{Carts}, making it harder to trace downstream bottlenecks. If either service slows down (Fig.~\ref{fig:bottleneckP90}), user-perceived performance may degrade, but autoscalers relying on call graphs might fail to attribute the root cause accurately.

Fig.~\ref{fig:bottleneckP90}B shows high latency at \texttt{User} but low latency at \texttt{Carts}. Scaling out only \texttt{User} can unintentionally overload \texttt{Carts}, causing SLO violations. This underscores the importance of considering the entire service chain in auto-scaling decisions.

\subsection{Security}
Auto-scaling must operate within safe resource boundaries to avoid destabilizing the system. Without proper constraints, scaling can lead to resource exhaustion or service disruption. This subsection outlines key security considerations, including resource limits and namespace isolation.

\vspace{0.2em}
\noindent\textbf{Resource Governance.} While readiness probes can suppress auto-scaler reactions to initialization spikes, they do not address the root cause: inadequate resource provisioning. In our experiments, the \texttt{Carts} service required nearly five minutes to initialize under default Kubernetes settings—significantly delaying responsiveness. A naïve solution might be to remove resource constraints entirely, but this introduces broader concerns around resource governance and overall cluster stability.

Kubernetes, by design, allows containers to operate without predefined CPU or memory constraints. While this flexibility supports rapid prototyping, it poses serious risks in multi-tenant or production environments. Below, we describe two specific issues observed in our deployment and how they relate to auto-scaling and system reliability.

\noindent\textbullet\ \textit{G9: Unbounded Resource Requests:}
Enforcing CPU and memory limits for pods or namespaces is essential to mitigate resource-exhaustion risks~\cite{XICommandments}. By default, Kubernetes allows unbounded CPU and memory access, enabling a misbehaving or compromised service to consume excessive resources. This can trigger cascading failures across the cluster—especially during auto-scaling events.

For example, if a resource-intensive service such as \texttt{Carts} is scaled out without limits, multiple replicas may launch simultaneously, each consuming significant CPU during JVM warm-up. This can starve co-located services, degrade overall performance, and increase pod eviction rates. Historical vulnerabilities have demonstrated how unbounded resource requests can be exploited to mount denial-of-service attacks~\cite{XICommandments}.

As illustrated in Fig.~\ref{fig:bottleneckP90}, improperly constrained services such as \texttt{Carts} or \texttt{User} can become performance bottlenecks, leading to latency spikes and SLO violations in downstream services like \texttt{Front-end}.

\noindent\textbullet\ \textit{G10: Lack of Namespace-Level Safeguards:}
Even when individual containers have resource limits, the lack of aggregate controls can lead to systemic overload. In large deployments, microservices are often grouped by team, feature, or environment within namespaces~\cite{Goldman}. Without namespace-level quotas, a single namespace can monopolize CPU or memory across the entire cluster, either due to misconfiguration or attack.

%\vspace{1.5em}

\section{Microservice Auto-Scaling Considerations}
Effective auto-scaling in microservices-based systems demands careful attention throughout the entire software lifecycle. Our evaluation revealed that many real-world scaling issues stem not from flaws in auto-scaling algorithms, but from shortcomings introduced during system design, implementation, or deployment. To systematically examine these issues, we adopt a three-phase lifecycle—\textit{Architecture}, \textit{Implementation}, and \textit{Deployment}—which reflects both empirical observations and foundational software engineering principles. This perspective helps identify when and where scaling challenges arise, enabling proactive mitigation rather than reactive tuning.

\vspace{0.3em}
\noindent\textbf{Architecture Phase.} Neglecting scalability during the architectural phase—such as unclear service dependencies or improper chaining—can significantly impair bottleneck detection. Without clear call relationships and runtime-aware observability (e.g., via service mesh traces), auto-scalers struggle to attribute resource pressure accurately, resulting in ineffective or misdirected scaling actions.

\vspace{0.3em}
\noindent\textbf{Implementation Phase.} Even well-architected systems can underperform if developers omit critical implementation elements. Missing init containers for boot-heavy tasks, misconfigured liveness/readiness probes, or the absence of application-level metrics (e.g., latency, custom error codes) all reduce the auto-scaler’s visibility, degrading both precision and responsiveness.

\vspace{0.3em}
\noindent\textbf{Deployment Phase.} Ineffective deployment configurations further undermine auto-scaling. Improperly tuned resource requests/limits, aggressive probe timeouts, or suboptimal autoscaler policies (e.g., thresholds, replica caps) can destabilize scaling, causing oscillations, delays, or even service outages.

\begin{table*}[ht]
\caption{Mapping of Challenges to the Three-Phase Lifecycle}
\label{tab:phase-mapping}
\centering
\renewcommand{\arraystretch}{1.2}
\begin{tabular}{@{}p{2.8cm} p{3cm} p{3.2cm} p{1cm} p{5.1cm}@{}}
\toprule
\textbf{Phase} & \textbf{Auto-scaling Challenge} & \textbf{Type} & \textbf{Gap ID} & \textbf{Key Consideration} \\
\midrule
\multirow{1}{*}{\textbf{Architecture}} 
  & Observability & Dependencies & G8 & Incomplete or Misleading Call Graphs \\
\midrule
\multirow{6}{*}{\textbf{Implementation}} 
  & Scalability & Heavy Services & G1 & Service Initialization Overhead \\
  & Scalability & Heavy Services & G2 & Scale-Out Resource Contention \\
  & Observability & Monitoring & G3 & Missing Application-Level Metrics \\
  & Observability & Failure Visibility & G6 & Error Masking in Service Chains \\
  & Observability & Failure Visibility & G7 & Lack of Downstream Error Metrics \\
\midrule
\multirow{3}{*}{\textbf{Deployment}} 
  & Observability & Readiness & G4 & Missing Readiness/Liveness configuration \\
    & Observability & Readiness & G5 & Misconfigured Readiness/Liveness Probes \\
  & Security & Resource Governance & G9 & Unbounded Resource Requests \\
  & Security & Resource Governance & G10 & Lack of Namespace-Level Safeguards \\
\bottomrule
\end{tabular}
\end{table*}

To contextualize these three phases within broader software engineering practice, we align them with widely adopted development methodologies: RUP~\cite{RUP}, DevOps~\cite{DevOps}, SAFe~\cite{SAFe}, Agile~\cite{Agile}, and the microservice migration model by~\cite{migration2025survey}. Table~\ref{tab:phase_mapping} shows how each methodology maps onto our proposed lifecycle model. This mapping highlights the generality of our framework and emphasizes when scaling-related decisions typically emerge during the development process.

%\addtocounter{table}{-2}
%\vspace{0.5em}
\begin{table}[H]
\caption{Mapping Software Development Methodologies to Proposed Phases (Listed alphabetically)}
\centering
\begin{tabular}{@{}llccc@{}}
\toprule
\textbf{Methodology} & \textbf{Stage} & \textbf{Archite.} & \textbf{Impleme.} & \textbf{Deploym.} \\
\midrule
\multirow{3}{*}{Agile~\cite{Agile}} 
  & Sprint 0        & \checkmark &          &          \\
  & Iteration       &           & \checkmark &         \\
  & Release         &           &          & \checkmark \\
\midrule
\multirow{3}{*}{DevOps~\cite{DevOps}} 
  & Plan/Design     & \checkmark &          &          \\
  & Develop/Test    &           & \checkmark &         \\
  & Release/Operate &           &          & \checkmark \\
\midrule
\multirow{5}{*}{\shortstack[l]{Migration \\Model~\cite{migration2025survey}}} 
  & Planning        & \checkmark &          &          \\
  & Analysis        & \checkmark &          &          \\
  & Design          & \checkmark &          &          \\
  & Execution       &           & \checkmark & \checkmark \\
  & Monitoring      &           &          & \checkmark \\
\midrule
\multirow{3}{*}{RUP~\cite{RUP}} 
  & Elaboration     & \checkmark &          &          \\
  & Construction    &           & \checkmark &         \\
  & Transition      &           &          & \checkmark \\
\midrule
\multirow{3}{*}{SAFe~\cite{SAFe}} 
  & PI Planning     & \checkmark &          &          \\
  & ART Execution   &           & \checkmark &         \\
  & Release Demand  &           &          & \checkmark \\
\bottomrule
\end{tabular}
\label{tab:phase_mapping}
\end{table}
\vspace{0.5em}
This lifecycle-oriented perspective enables practitioners and researchers to classify and address auto-scaling challenges in a structured and proactive manner. In the following sections, we examine issues observed in each phase using real-world microservice benchmarks.

Table~\ref{tab:phase-mapping} presents a consolidated summary of the key auto-scaling considerations identified through our evaluation. Each row corresponds to a specific issue, organized under a high-level theme and subcategory, and mapped to the relevant phase of the microservice lifecycle where design or mitigation effort is needed. This structured view supports early identification of scaling concerns and facilitates the design of systems that remain robust under dynamic workload conditions.

\section{Evaluation Setup and Methodology}\label{sec:evaluation}
To assess the practical impact of the identified auto-scaling considerations, we conducted a series of experiments using the Sock-Shop benchmark. Sock-Shop was selected as the primary evaluation target because it captures a wide spectrum of real-world architectural and operational characteristics, and it exhibits multiple auto-scaling pitfalls discussed in Section~\ref{sec:challenges}.

%\addtocounter{table}{+1}
%\vspace{0.5em}
%\subsection*{Experimental Environment}
\vspace{0.3em}
\noindent\textbf{Experimental Environment.}
Experiments were conducted on a high-performance server equipped with dual AMD EPYC 7742 processors (256 vCPUs) and 1~TB RAM. This hardware was partitioned into virtual machines to host a Kubernetes control plane and two worker nodes. Microservices were deployed using standard Kubernetes manifests, and each auto-scaler was configured following its original documentation.

%\vspace{0.5em}

%\subsection*{Workload Generation}
\vspace{0.3em}
\noindent\textbf{Workload Generation.}
To simulate dynamic, real-world traffic patterns, we employed the \texttt{LOCUST.io}\cite{locust} load testing framework in combination with the WorldCup98 dataset\cite{WorldCup}, which captures both peak and off-peak demand fluctuations. The synthetic workload primarily targeted the Sock-Shop \texttt{/login} service. %, which exercises key backend components: \texttt{Front-end}, \texttt{User}, and \texttt{Carts}.

%\vspace{0.5em}
%\subsection*{Evaluated Auto-Scaling Methods}
\vspace{0.3em}
\noindent\textbf{Evaluated Auto-Scaling Methods.}
We evaluated six representative auto-scaling methods spanning threshold-based, control-theoretic, learning-based, and topology-aware strategies:

\begin{itemize}[leftmargin=0.3cm]
  \item \textbf{KHPA} — Kubernetes' default autoscaler that adjusts replicas based on static CPU thresholds.
  \item \textbf{HEAT} — Combines resource thresholds with linear regression to predict short-term load~\cite{HEAT}.
  \item \textbf{SHOWAR} — Applies a PID controller to stabilize resource utilization~\cite{SHOWAR}.
  \item \textbf{Fixed-PID} — Enhances PID control using an offline-trained neural model to adaptively tune gain parameters~\cite{FPID}.
  \item \textbf{MicroScaler} — Uses Bayesian optimization to search for near-optimal replica counts~\cite{Microscaler}.
  %uses Bayesian optimization to treat scaling as a black-box search problem \cite{Microscaler}, offering near-optimal results but with higher computational overhead.
  \item \textbf{PBScaler} — Leverages dynamic call graphs and genetic search to locate and scale only bottlenecked services~\cite{PBScaler}.
  %combines runtime call-graph analysis with genetic optimization to locate bottlenecks and scale only the affected services \cite{PBScaler}, representing an advanced topology-aware strategy.
  %
\end{itemize}

\vspace{0.5em}
%\subsection*{Challenge Mitigation Summary}
\noindent\textbf{Challenge Mitigation Summary.}\label{sec:mitigation} To ensure a fair evaluation of the auto-scaling methods, we addressed as many of the previously identified Sock-Shop issues as feasible.

\begin{table}[ht]
\caption{Summary of Problem Remediation in Sock-Shop}
\label{tab:problem-fixes}
\centering
\renewcommand{\arraystretch}{1.15}
\begin{tabular}{@{}p{0.4cm}p{7.9cm}@{}}
\toprule
\textbf{ID} & \textbf{Remediation Strategy} \\
\midrule
G1 & We were unable to use \texttt{initContainers} to isolate JVM startup. Instead, we manually modified the Linux cgroup configuration to temporarily allocate more CPU during the boot phase. \\
\addlinespace[1.5pt]
G2 & We modified the auto-scaling method source code to enforce a cap of 5 replicas for the \texttt{Carts} microservice to avoid resource contention during scale-out events. \\
\addlinespace[1.5pt]
G3 & The \texttt{Carts} service did not expose health metrics. We modified its YAML configuration to inject JVM options that enable a health check endpoint for readiness probing. \\
\addlinespace[1.5pt]
G4 & After enabling the health endpoint, we configured readiness and liveness probes in the deployment YAML for \texttt{Carts}. \\
\addlinespace[1.5pt]
G5 & We tuned the probe timeouts and initial delays according to our infrastructure performance to avoid premature restarts. \\
\addlinespace[1.5pt]
G6 & Due to frontend exception masking and compilation challenges, we instead monitored \texttt{Carts} logs for failure patterns. When failures were detected, we restarted the \texttt{Carts} pod manually. \\
\addlinespace[1.5pt]
G7 & Not applicable. \\
\addlinespace[1.5pt]
G8 & We attempted to modify the source code to reflect a chained invocation pattern, but encountered compilation issues. As a workaround, we hard-coded the call graph for the \texttt{/login} service directly into the auto-scaling method. \\
\addlinespace[1.5pt]
G9 & Not applicable—CPU and memory limits were already defined for all services. \\

G10 & We enforced namespace-level quotas and replica caps to prevent uncontrolled replica creation. \\
\bottomrule
\end{tabular}
\end{table}
%\vspace{-0.5em}

\begin{figure*}[ht]
  \centering
  \includegraphics[width=\textwidth]{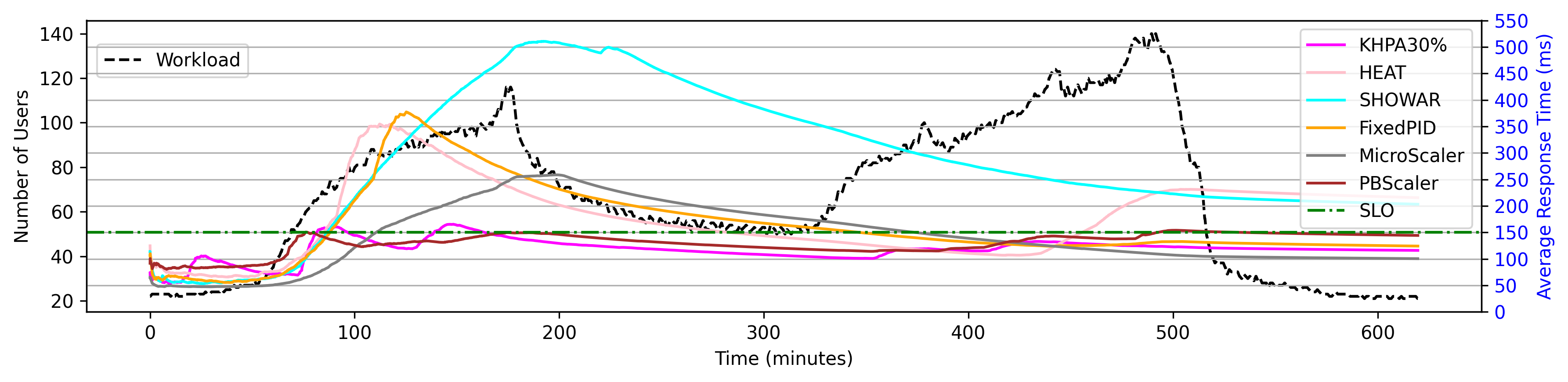}
  \vspace{-1.5em}
  \caption{Average end-to-end response time over 10 hours - SLO threshold: 150~ms. %; Resource configuration: Requests: 100m, Limits: 300m.
  }
  \vspace{-1em}
  \label{fig:resp_time_plot}
\end{figure*}

\begin{figure*}[ht]
  \centering
  \includegraphics[width=\textwidth]{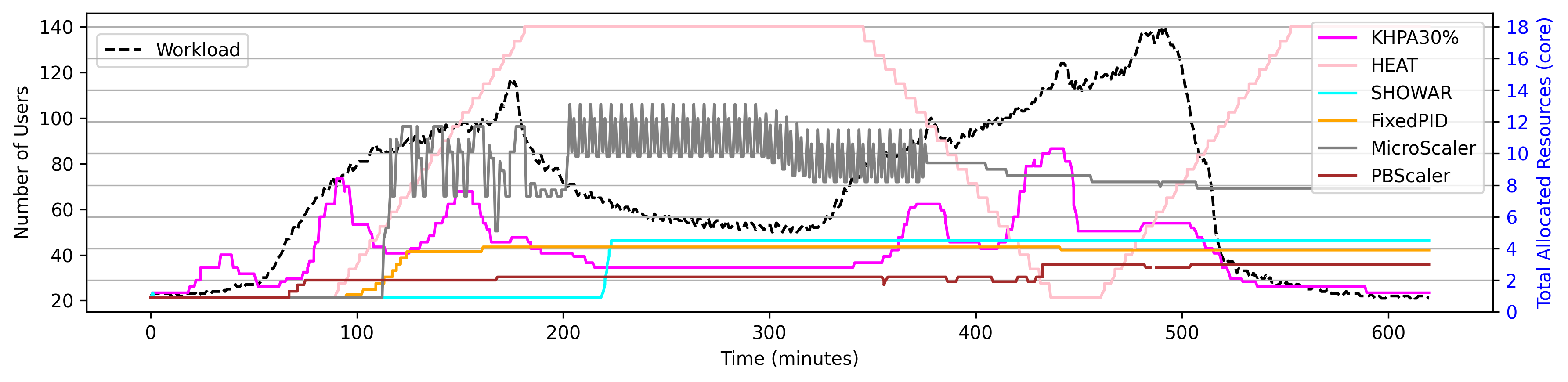}
  \vspace{-1.5em}
  \caption{Total CPU core-minutes used by \texttt{Front-end}, \texttt{User}, and \texttt{Carts} - Max pods for each microservice: 20.
  % - Max pods: 20; Resource configuration: Requests: 100m, Limit: 300m
  }
  \vspace{-1em}
  \label{fig:resource_usage_plot}
\end{figure*}

Specifically, we categorized the ten gaps as follows: G1, G3, G4, G5, and G6 were essential for enabling auto-scaling on the \texttt{/login} path and were addressed. G9 and G7 were not applicable in this context, while G2, G8, and G10 were incorporated as general best practices. We made targeted modifications to improve probe configurations, resource governance, and application observability. Table~\ref{tab:problem-fixes} summarizes the applied remediations and their scope.

%To ensure a fair evaluation of the auto-scaling methods, we addressed as many of the previously identified Sock-Shop issues as feasible. We made several modifications to improve probe configurations, resource governance, and application observability. Table~\ref{tab:problem-fixes} summarizes the applied remediations and their scope.

\section{Experimental Results}\label{sec:results}
%We evaluated the impact of autoscaling considerations by deploying Sock-Shop’s \texttt{/login} service—comprising \texttt{Front-end}, \texttt{User}, and \texttt{Carts}—under six different autoscaling methods. Each microservice was capped at 20 pods, and Kubernetes resource settings were: \texttt{Front-end} (200m/300m), \texttt{User} (200m/300m), and \texttt{Carts} (400m/400m) for Requests/Limits. We generated load using LOCUST with the WorldCup98 traffic trace over a 10-hour period (600 minutes). The two primary evaluation metrics were: (1) mean end-to-end response time as observed by users, and (2) total CPU usage (in core-minutes) aggregated across the three services.

We evaluated the impact of auto-scaling considerations by deploying Sock-Shop’s \texttt{/login} service—comprising the \texttt{Front-end}, \texttt{User}, and \texttt{Carts} microservices—under six different auto-scaling methods. Each microservice was capped at a maximum of 20 pods, with Kubernetes resource configurations as follows: \texttt{Front-end} (200m/300m), \texttt{User} (200m/300m), and \texttt{Carts} (400m/400m) for Requests/Limits, respectively. Load was generated using \texttt{LOCUST}, driven by the WorldCup98 traffic trace over a 10-hour (600-minute) period. The two primary evaluation metrics were: (1) mean end-to-end response time as observed by users, and (2) total CPU usage (in core-minutes) aggregated across the three services.

\vspace{0.3em}
\noindent \textbf{Assumption.} Baseline (unmitigated) results are omitted, as key services were not autoscaling-ready without the mitigations described in Section~\ref{sec:mitigation}.

%To clarify the role of the ten identified gaps (G1–G10) in our Sock-Shop experiment, we categorized them based on their applicability. Gaps G1, G3, G4, G5, and G6 were necessary to enable auto-scaling on the \texttt{/login} path. G2, G7, and G8 were not applicable in this scenario. G9 and G10, while not essential for enabling auto-scaling, are included as recommended practices to improve autoscaler stability and resource governance. Note that not all gaps are relevant to every benchmark microservice, but collectively they capture common patterns observed in practice.

Fig.~\ref{fig:resp_time_plot} shows that PBScaler consistently achieved the lowest mean response time, satisfying the 150ms SLO. In contrast, HEAT and SHOWAR experienced frequent violations despite employing predictive or feedback-based logic. Fig.\ref{fig:resource_usage_plot} presents the total CPU usage across services. PBScaler again outperformed other methods, minimizing resource consumption by scaling only true bottlenecks. HEAT, despite aggressive overprovisioning, failed to maintain latency targets. KHPA exhibited lower CPU usage but suffered from frequent SLO violations due to its reliance on static thresholds.

%\vspace{-0.5em}
\begin{table}[ht]
\caption{Aggregate SLO Violations and CPU Usage}
\centering
\renewcommand{\arraystretch}{1.2}
\begin{tabular}{lcc}
\toprule
\textbf{Method} & \textbf{SLO Violations} & \textbf{CPU Core-Minutes} \\
\midrule
KHPA              & 1,563     & 11,754  \\
HEAT              & 134,754   & 34,288  \\
SHOWAR            & 450,052   & 9,966   \\
Fixed-PID         & 98,320    & 10,630  \\
MicroScaler       & 67,423    & 23,668  \\
PBScaler          & 387       & 6,928  \\
\bottomrule
\end{tabular}
\label{tab:results}
\end{table}
%\vspace{-0.5em}

As summarized in Table~\ref{tab:results}, PBScaler demonstrated superior performance across both response time and CPU efficiency metrics. Its integration of topology awareness and runtime observability enabled more accurate and targeted scaling decisions. These results reinforce the insight that successful auto-scaling depends not only on algorithmic sophistication, but also on addressing the architectural, implementation, and deployment-level challenges outlined in Section~\ref{sec:challenges}. In contrast, simpler approaches such as KHPA and HEAT—while easier to configure—struggled to adapt to real-world workload dynamics.

\section{Lessons Learned}\label{sec:lessons}
Our empirical evaluation of six auto-scaling methods on the Sock-Shop benchmark—conducted while incrementally resolving real-world deployment and observability issues—revealed several key insights that inform both the design and operational use of microservice auto-scalers:

\noindent \textbf{Auto-scaling effectiveness goes beyond algorithm design.} Even advanced methods such as PBScaler depend on accurate service call graphs, observable error signals, and comprehensive metric instrumentation. Without addressing architectural and implementation-level deficiencies, these systems often fail to outperform simpler baselines.

\noindent\textbf{Heavy services create systemic stress.} Resource-intensive microservices with long initialization times can overload both the autoscaler and the cluster. Mitigation requires architectural forethought and carefully managed scale-out strategies.

\noindent\textbf{Failure visibility is essential.} When downstream errors are masked, auto-scalers lack the signals needed to respond to service degradation. In our experiments, exposing explicit error metrics and surfacing failures through logs were essential to enable meaningful scaling responses.

\noindent\textbf{Probes must be properly configured.} Misconfigured readiness and liveness probes can destabilize service behavior, particularly under load or during cold starts. Accurate probe tuning, aligned with infrastructure performance, is crucial for reliable scaling.

\noindent\textbf{Auto-scaling is a lifecycle-wide concern.} Scalability issues arise during architecture, implementation, and deployment. Treating auto-scaling as a runtime-only task overlooks key factors that determine its effectiveness.

\smallskip
\noindent In summary, autoscaler performance depends not only on algorithm quality, but on the readiness of the underlying system to support scalable behavior throughout its lifecycle.
%\noindent In summary, the effectiveness of auto-scaling is shaped not only by the sophistication of the algorithm, but also by the system’s architectural readiness to support scalable behavior throughout its lifecycle.

%\noindent\textbf{A clean system state is required for fair evaluation.} Comparing autoscaling strategies without addressing observability and deployment issues leads to misleading conclusions. Many metrics only became meaningful after ensuring system health and metric propagation.

\section{Related Works}\label{sec:background}
This section reviews prior work on microservice architecture and migration, auto-scaling techniques, and security risks related to scaling. We cover decomposition strategies, spatial-temporal scaling approaches, and misconfiguration vulnerabilities in autoscaler settings.

\noindent\textbf{Microservice Architecture.}
Designing microservices—either from scratch or by migrating from monolithic systems—has been widely studied. The literature outlines best practices for decomposition, communication, and deployment to improve maintainability and scalability. Migration-specific works, such as Saucedo \textit{et al.}~\cite{migration2025survey}, classify key phases including planning, decomposition, and post-deployment verification. Studies by Francesco \textit{et al.}~\cite{francesco2019architecting} and Razzaq \textit{et al.}~\cite{razzaq2023systematic} highlight challenges such as data consistency, inter-service coordination, and organizational readiness, while model-driven recovery approaches like MiSAR~\cite{misar2025} aim to address architectural clarity and consistency during migration.

Empirical analyses have also examined decomposition strategies~\cite{fritzsch2019microservices, fritzsch2019classification}, tool support~\cite{microdecomposition}, and post-migration tradeoffs~\cite{soldani2018pains,wolfart2021modernizing}. While these works provide valuable guidance for building microservices, they often focus on maintainability and modularity, with limited emphasis on runtime operational concerns like auto-scaling compatibility.

%\subsection*{Auto-Scaling in Microservices}
\noindent\textbf{Auto-Scaling in Microservices.}
While microservices are built for scalability, this requires both elastic infrastructure and tailored auto-scaling. Default solutions like AWS Auto Scaling~\cite{Aws} and Kubernetes HPA often overlook fine-grained service interactions. Effective scaling demands spatiotemporal awareness of workloads, including request bursts, service dependencies, and performance metrics~\cite{Dashtbani2025}.

Temporal-aware solutions include ARAScaler~\cite{ARAScaler}, which adapts resource scaling with ETimeMixer; PBScaler~\cite{PBScaler}, which uses TopoRank to find bottlenecks; DeepScaler~\cite{DeepScaler}, which applies attention-based GCNs to coordinate scaling, reducing SLA violations by 41\%; and Microscaler~\cite{Microscaler}, which uses a Service Power metric and online learning to cut response times by 15\% and failures by 24\%.

Spatial features involve service roles, execution timing, and dependencies. STaleX~\cite{Dashtbani2025} uses PID controllers adjusted in real time based on spatial and temporal inputs, reducing resource use by 26.9\%. DCScaler~\cite{DCScaler} forecasts service demand using call graphs to coordinate distributed scaling. STAAF~\cite{STAAF} models spatial–temporal dependencies to maintain SLA compliance. MarVeLScaler~\cite{MarVeLScaler}, originally for MapReduce, applies multi-view deep learning to predict cluster sizes.% with 98.4\% accuracy and reduce costs by 30.8%.

%\subsection*{Security Considerations in Auto-Scaling}
\noindent\textbf{Security Considerations in Auto-Scaling.}
Auto-scaling introduces security risks when resource policies are misconfigured or missing. Ben David~\cite{yoyoattack} demonstrates the \emph{YoYo attack}, where attackers manipulate load to trigger excessive scale events, leading to resource exhaustion.

Kubernetes manifests are prone to misconfigurations that compromise autoscaler safety. Studies by Shamim \textit{et al.}~\cite{XICommandments,Manifests,IaCodemit} reveal recurring issues—such as missing CPU/memory limits or replica caps—that can allow a single service to monopolize resources.

We also reviewed CVE datasets~\cite{nistNVD,announce}, identifying autoscaling-related risks such as uncontrolled replica creation and denial-of-service vectors. These findings highlight the importance of integrating resource quotas and scaling safeguards as part of secure autoscaler deployment.

\vspace{0.5em}
\noindent
Despite progress in autoscaling research, little attention has been paid to the readiness of benchmark microservices. This paper fills that gap by identifying practical design issues that hinder autoscaler effectiveness and demonstrating how addressing them improves real-world performance.

\section{Conclusion and Future Work}
\label{sec:conclusion}
This paper presented a systematic analysis of auto-scaling challenges in microservice-based systems, grounded in practical issues encountered during the deployment and evaluation of widely used microservice benchmarks. We proposed a three-phase lifecycle—\textit{Architecture}, \textit{Implementation}, and \textit{Deployment}—to organize these challenges across the software development lifecycle.

Using Sock-Shop as a case study, we demonstrated how real-world design flaws—such as incomplete call graphs, error masking, and misconfigured probes—impair the effectiveness of autoscaling methods. Our experiments showed that addressing these issues enables advanced techniques like PBScaler to significantly outperform reactive or threshold-based strategies in both SLO compliance and resource efficiency. In summary, auto-scaling is not solely an algorithmic problem; it requires coordinated consideration of system architecture, observability, and deployment configurations.

%\textbf{Future Work.}
In future work, we plan to extend our evaluation to additional microservice benchmarks (e.g., TrainTicket, Online Boutique); refine the identified challenges into a reusable checklist for system architects; integrate observability validation into CI/CD pipelines to detect auto-scaling anti-patterns early; and investigate automated remediation techniques (e.g., probe tuning) to minimize the manual effort required to prepare services for intelligent scaling.

\end{document}